\documentclass[aps,pra,twocolumn,superscriptaddress]{revtex4}
\usepackage{graphicx}
\usepackage{hyperref}
\usepackage{xcolor}
\usepackage{amsmath}
\usepackage{amsfonts}
\usepackage{amssymb}

\begin{document}
\title{Electromagnetically Induced Transparency in $\Lambda$-systems of $^{87}Rb$ atom in magnetic field}
\author{Charu Mishra}
\email{charumishra@rrcat.gov.in}
\affiliation{Laser Physics Applications section, Raja Ramanna Center for Advanced Technology, Indore 452013, India}
\affiliation{Homi Bhabha National Institute, Mumbai-400094, India}
\author{A. Chakraborty}
\affiliation{Laser Physics Applications section, Raja Ramanna Center for Advanced Technology, Indore 452013, India}
\affiliation{Homi Bhabha National Institute, Mumbai-400094, India}
\author{A. Srivastava}
\affiliation{Laser Physics Applications section, Raja Ramanna Center for Advanced Technology, Indore 452013, India}
\author{S. K. Tiwari}
\affiliation{Laser Physics Applications section, Raja Ramanna Center for Advanced Technology, Indore 452013, India}
\author{S. P. Ram}
\affiliation{Laser Physics Applications section, Raja Ramanna Center for Advanced Technology, Indore 452013, India}
\author{V. B. Tiwari}
\affiliation{Laser Physics Applications section, Raja Ramanna Center for Advanced Technology, Indore 452013, India}
\affiliation{Homi Bhabha National Institute, Mumbai-400094, India}
\author{S. R. Mishra}
\affiliation{Laser Physics Applications section, Raja Ramanna Center for Advanced Technology, Indore 452013, India}
\affiliation{Homi Bhabha National Institute, Mumbai-400094, India}

\begin{abstract}
The electromagnetically induced transparency (EIT) observations in two $\Lambda$-systems of $^{87}Rb$ atom, $|5^{2}S_{1/2} F=1\rangle \rightarrow |5^{2}P_{3/2} F'=1\rangle \leftarrow |5^{2}S_{1/2} F=2\rangle$  and $|5^{2}S_{1/2} F=1\rangle \rightarrow |5^{2}P_{3/2} F'=2\rangle \leftarrow |5^{2}S_{1/2} F=2\rangle$, have been investigated in detail and the results are found consistent with our proposed theoretical models. The second $\Lambda$-system provides EIT signal with higher magnitude than the first system, both in absence and in presence of an applied magnetic field. The observed steeper slope of the EIT signal in presence of the magnetic field can enable one to achieve tight frequency locking of lasers using these EIT signals. 
\end{abstract}

\maketitle

\section{Introduction}
The electromagnetically induced transparency (EIT), a quantum interference  phenomenon, is related to increase in the transmission of a weak probe beam in presence of a strong coupling beam \cite{Harris:1990,Harris:1st}.  Since the conception of EIT, several applications of this phenomenon have  attracted attention of researchers which include applications in optical communication network \cite{Ma:2017}, optical switching devices \cite{Clarke:12:2001,Ray:2013}, manipulation of group velocity of light and light storage \cite{Kash:1999,Kocharovskaya:2001,Karpa:2006,Goldfarb:2008,Alotaibi:2015}, tight laser frequency locking \cite{Bell:2007}, high resolution spectroscopy \cite{Krishna:2005,Kale:2015}, atomic clocks \cite{Guidry:2017}, quantum information processing \cite{Beausoleil:eit,Beausoleil:cpt}, highly sensitive magnetometery  \cite{Lee:1998,Fleischhauer:1994,Yudin:2010,Cox:2011,Margalit:2013,Cheng:2017}, velocimetry \cite{Santos:2006}, etc.

The simple schemes to obtain EIT involve three atomic levels, in which the weak probe laser beam is resonant to one pair of energy levels and the strong coupling beam is resonant to another pair of energy levels such that one level is common to probe and coupling both. These configurations are known as ladder \cite{Moon:2005}, lambda \cite{Li:1995} and vee \cite{Kang:2014} systems. Albeit, detailed studies of these configurations already exist in the literature \cite{Fulton:116:1995,Fulton:52:1995,Clarke:64:2001,Yan:2001,VBT:2010,Xiaojun:2016}, proper comparative studies of a particular atomic system are hard to find to decide the better system for a specific application.

In this article, our objective is to provide a detailed account of the EIT phenomenon for the two possible $\Lambda$-systems of $D_{2}$ line of $^{87}Rb$ atom by varying the coupling beam and magnetic field strength to find the better system for EIT based laser locking applications. The two $\Lambda$-systems, denoted hereafter as system (A) and system (B) respectively, are formed with the atomic levels $|5^{2}S_{1/2} F=1\rangle \rightarrow |5^{2}P_{3/2} F'=1\rangle \leftarrow |5^{2}S_{1/2} F=2\rangle$  and $|5^{2}S_{1/2} F=1\rangle \rightarrow |5^{2}P_{3/2} F'=2\rangle \leftarrow |5^{2}S_{1/2} F=2\rangle$. Both these $\Lambda$-systems have been investigated in absence and in presence of an applied magnetic field. These two $\Lambda$-systems differ in terms of the strength of transitions and the proximity of other energy levels surrounding the excited state. In presence of magnetic field, the Zeeman splitting of levels further differentiates the two $\Lambda$-systems. The following are highlights of the experimental observations which have been found consistent with our  theoretical models. In absence of a magnetic field for a particular coupling beam power, system (B) provides the EIT signal of higher magnitude and more symmetric line shape than the system (A). The presence of magnetic field splits the single EIT peak into three peaks where the central peak of system (B) again exhibit better strength. At a higher coupling beam power, this central EIT peak possess steeper slope than that of the peak observed in absence of the magnetic field. These signals with steeper slope can be extremely useful for tight frequency locking of lasers. Also, the linear dependence of the separation between the split peaks on the magnetic field strength can be exploited to lock a laser frequency at a frequency detuning controlled by the magnetic field. These studies may be performed in near future.

The article is organized as follows. Section \ref{sec:expt} describes the experimental setup. Our results and discussion are presented in section \ref{sec:result}. Finally, the conclusion of our work is given in section \ref{sec:concl}.

\section{Experimental Setup}
\label{sec:expt}

\begin{figure}[t]
\centering
\includegraphics[width=8.5 cm]{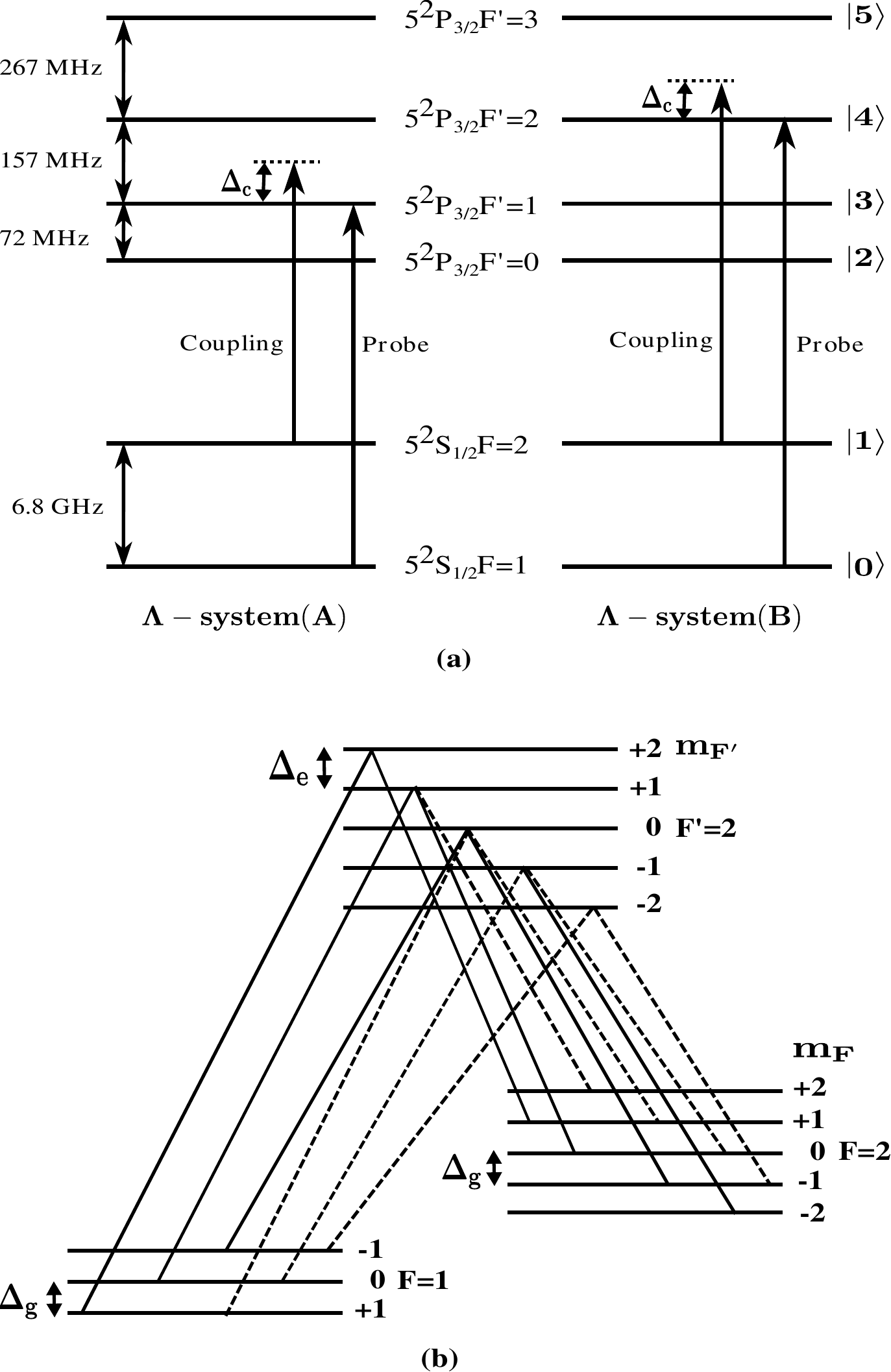}
\caption{ (a) The relevant energy levels of $^{87}Rb$ atom showing two $\Lambda$-systems investigated in the present work.  $\Delta_{c}$ is the detuning of the coupling beam from respective transition resonance. (b) $\Lambda$-subsystems within the system (B) in presence of a longitudinal magnetic field. $\Delta_{g}$ and $\Delta_{e}$ are energy separations between magnetic sublevels of ground and excited states respectively. Solid and dashed lines connecting levels represent transitions involving $\sigma^{+}$ and $\sigma^{-}$ polarizations.}
\label{lvl}
\end{figure}

\begin{figure}[t]
 \centering
 \includegraphics[width=8.5cm]{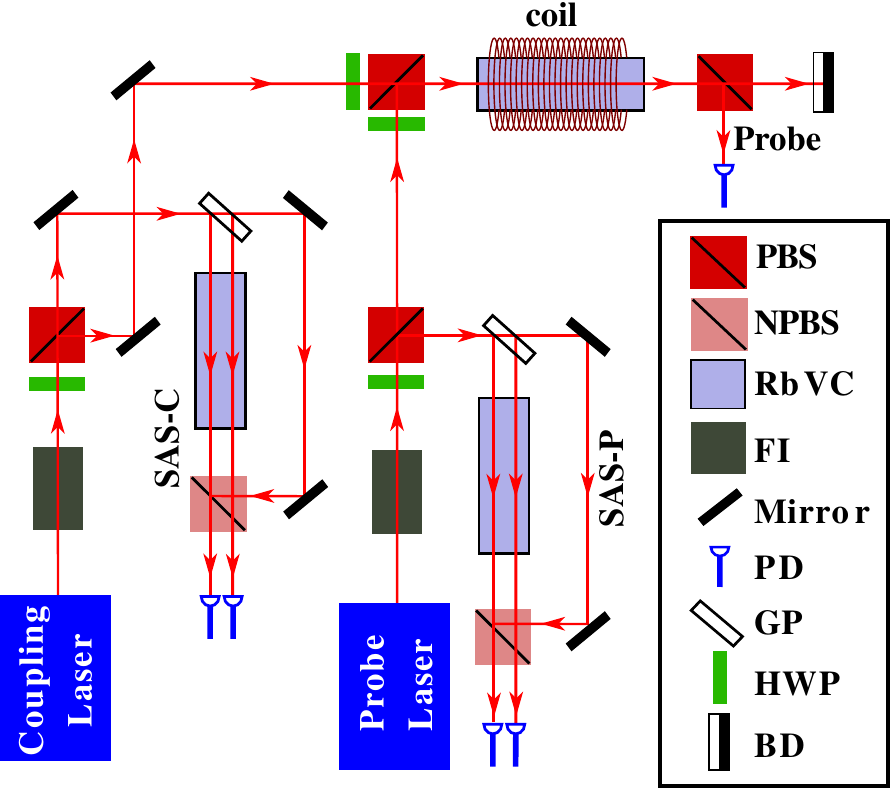}
 \caption{(Color online) Schematic diagram of the experimental setup. PBS: polarising beam splitter; NPBS: non-polarising beam splitter; RbVC: Rubidium vapor cell; FI: Faraday isolator, PD: photodiode; GP: glass plate; HWP: half-wave plate; BD: beam dump; SAS-P and SAS-C: saturated absorption spectroscopy for the probe and coupling beams respectively.}
 \label{setup}
\end{figure} 

The relevant energy levels of $^{87}Rb$ atom for the systems (A) and (B) are shown in Figure \ref{lvl}(a). The two $\Lambda$-systems were formed using a weak probe beam and a strong coupling beam connecting the two ground states with a common excited state. The probe beam frequency was fixed at the transition peak $|5^{2}S_{1/2} F=1\rangle \rightarrow |5^{2}P_{3/2} F'=1\rangle$ for system (A) and $|5^{2}S_{1/2} F=1\rangle \rightarrow |5^{2}P_{3/2} F'=2\rangle$ for system (B), while the coupling beam frequency was scanned across the transitions $|5^{2}S_{1/2} F=2\rangle \rightarrow |5^{2}P_{3/2} F'=1, 2,3\rangle$ for both the systems. The resonant probe beam gives a flat transmission signal over which the effect of the scanned coupling beam can be detected easily. The transition peaks were identified using the saturation absorption spectroscopy (SAS) technique. 

The schematic diagram of the experimental setup is shown in Figure \ref{setup}. The probe and coupling beams were derived from two independent external cavity diode lasers (TA-Pro and DL-100, TOPTICA, Germany) operating at 780 nm wavelength having spectral linewidths less than 1 MHz. The $1/e^{2}$ radii of the probe and coupling beams were 1.36 mm and 2.04 mm respectively. In order to make the systems Doppler free and insensitive to the atomic velocity, the probe and coupling beams were aligned in co-propagating geometry and  passed through a 50 mm long $Rb$ vapor cell \cite{carvalho}.  The pressure inside this vapor cell was $\sim 3.6 \times 10^{-7}$ Torr corresponding to a number density of $Rb$ atoms $\sim 1.2 \times 10^{10}\ cm^{-3}$. The magnetic field was applied using a current carrying solenoid wrapped over the Rb vapor cell. The length of the solenoid was much longer than the cell length to ensure the homogeneity of the applied magnetic field. Both the coil and the vapor cell were kept inside two layers of $\mu$-metal sheet to shield the cell from the stray magnetic field. The polarizations of the coupling and probe beams were kept linear but mutually orthogonal to each other. The power in the probe and coupling beams ($P_{p}$ and $P_{c}$ respectively) were controlled using a combination of a half-waveplate (HWP) and a polarizing beam splitter (PBS) in the beam paths. The transmitted probe and coupling beams were separated after the vapor cell using a PBS as shown in Figure \ref{setup}. The transmitted probe beam was collected on a photo-diode (connected to an oscilloscope) to measure the transmitted probe signal as a function of coupling beam detuning.

\section{Results and Discussion}\label{sec:result}

\subsection{Experimental results}\label{sec:expt_result}

\begin{figure}[t]
\centering
\includegraphics[width=8.5cm]{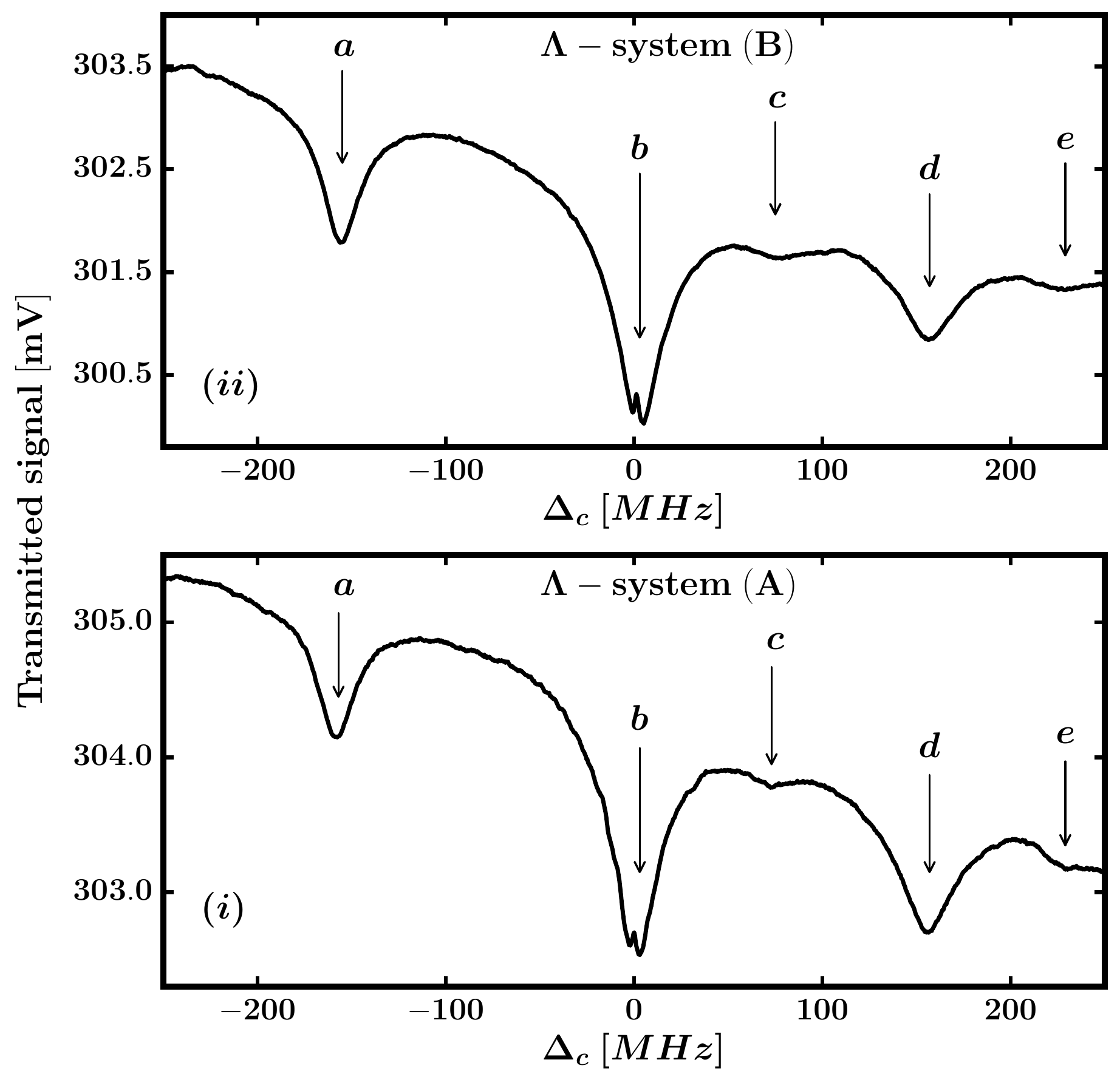}
\caption{ The transmitted probe signal as a function of coupling beam detuning. The spectra for both the systems show five absorption dips a, b, c, d and e, which are due to velocity selective optical pumping. Here only dip b shows a narrow EIT peak at zero detuning due to fulfilment of two photon resonance condition. The probe power is $P_{p}=0.08$ mW and the coupling beam power is $P_{c}=1$ mW}.
\label{vsop}
\end{figure}

In the experiments, the power of the probe beam was fixed at $P_{p}=0.08$ mW (Rabi frequency $\Omega_{p}=2\pi \times 3.3$ MHz for both the systems (A) and (B)) and the coupling beam power was varied in a wide range from $P_{c}=0.1$ mW to $P_{c}=20$ mW (resulting in a variation of the  coupling Rabi frequency from $\Omega_{c}=2\pi \times 1.1$ MHz to $\Omega_{c}=2\pi \times 15.5$ MHz for system (A) and $\Omega_{c}=2\pi \times 2.4$ MHz to $\Omega_{c}=2\pi \times 34.7$ MHz for system (B)). The Rabi frequency was calculated using the expression $\Omega=\Gamma \sqrt{\frac{I}{2I_{sat}}}$, where $I$ is the beam intensity, $\Gamma$ is the natural linewidth ($2\pi \times 6$ MHz) and $I_{sat}=\frac{c \epsilon_{0} \Gamma^{2} \hbar^{2}}{4d^{2}}$ is saturation intensity. The dipole moment $d=\mu_{ij} \mu_{0}$ with $\mu_{0}=3.58\times10^{-29}$ C-m and $\mu_{ij}$ is dipole transition matrix element between states $i$ and $j$. Here, $c, \epsilon_{0}$ and $\hbar$ are speed of light, vacuum permittivity and  reduced Planck's constant with values $3 \times10^8$ m/s, $8.85 \times10^{-12}$ F/m and $1.05 \times 10^{-34}$ J-s respectively.  Since both the systems (A) and (B) have equal $\mu_{ij}$ corresponding to the probe transition, the probe beam Rabi frequency $\Omega_{p}$ is equal for both the systems. Figure \ref{vsop} shows the recorded probe signals in which there are five absorption dips (a, b, c, d and e) for both the $\Lambda$-systems. These spectral features are known as velocity selective optical pumping (VSOP) absorption dips \cite{Iftiquar:2009,Chakrabarti:2005,Hossain:2011}. Among these absorption dips, only dip b exhibits an EIT peak due to the fulfillment of the two-photon resonance condition. In the present work, further investigation of this EIT peak in both the $\Lambda$-systems has been carried out by varying different experimental parameters.

\begin{figure}[t]
\centering
\includegraphics[width=8.5cm]{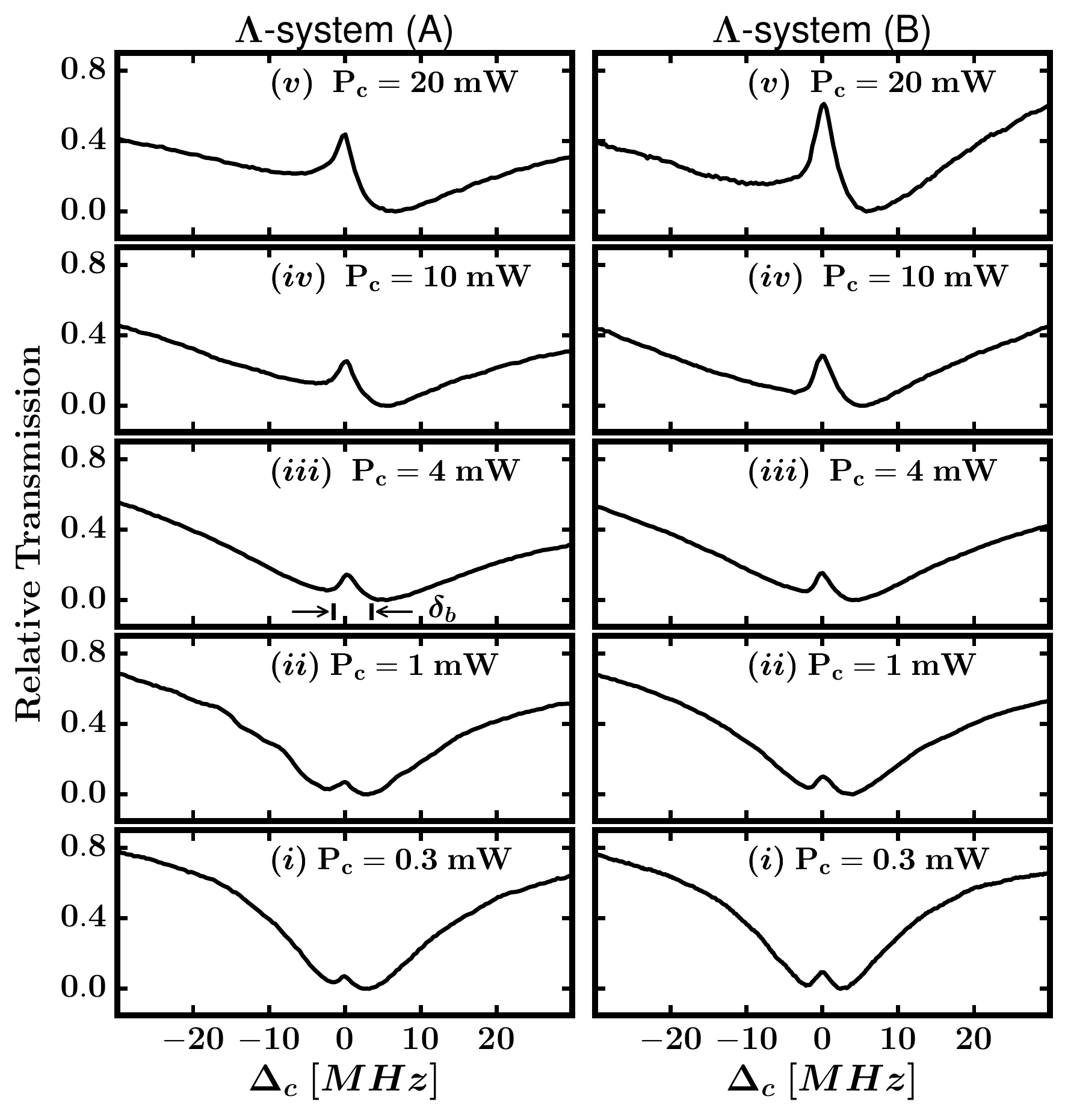}
\caption{ The relative transmission as a function of coupling beam detuning for both the $\Lambda$-systems (A) and (B), for different coupling beam power and the fixed probe beam power $P_{p}= 0.08$ mW. Here $\delta_{b}$ represents separation between minima of the EIT peak}.
\label{spectrum}
\end{figure}

The dependence of the EIT signal on the coupling beam power for both the $\Lambda$-systems was observed and the corresponding EIT signals are shown in Figure \ref{spectrum}. The relative transmission in the figure is defined as the ratio of the signal height (with respect to the VSOP minimum) to the total depth of corresponding VSOP absorption dip. This chosen scale provides the direct measure of recovery of the transmission due to EIT effect against the VSOP absorption, without involving the absolute value of probe signal voltage. This relative transmission ($T_R$) can be written in terms of regular transmission ($T$) and transmission at VSOP dip  ($T_{VSOP}$) as, $T_R=\frac{T-T_{VSOP}}{1-T_{VSOP}}$.

The important features of these observed EIT signals are as follows: (i) both strength and linewidth of the EIT signals increase with increase in the coupling beam power, (ii) the asymmetry in the line shape increases with the coupling power, (iii) the linewidth of the signal remains sub-natural even at higher coupling power, and (iv) the system (B) gives more symmetric and higher strength EIT signal than the system (A). Here, the linewidth is defined by parameter $\delta_b$ as shown in Figure \ref{spectrum} (\textit{iii}). Since the lineshape of the observed EIT signal is asymmetric, so measurement of the full width at half maximum (FWHM) of the EIT signal is difficult. Therefore, the separation between two minima in the EIT signal (\textit{i.e.} $\delta_b$) is considered as linewidth of the EIT signal. The actual FWHM of the EIT peak could be approximately the half of $\delta_b$. The strength of the EIT peak implies here the height of the EIT peak. The asymmetry in the signal is defined as difference between two minima values surrounding the EIT peak. 

The asymmetry in line shape of the EIT signal may be due to the presence of other nearby excited states, as reported earlier \citep{Bharti:2012,Chen:2013}. A more asymmetric EIT signal in the $\Lambda$-system (A) than that in the system (B) could possibly be due to this effect, as excited state in the system (A) has more closely spaced nearby levels than the system (B). The linewidth increases with the increase in the coupling beam power, while both the systems exhibit nearly same linewidth for a given coupling power. But, the system (B) shows stronger and more symmetric EIT signal than the system (A) as shown in Figure \ref{spectrum}. These results suggest that the $\Lambda$-system (B) can be more suitable for practical applications than the $\Lambda$-system (A).

\begin{figure}[t]
\centering
\includegraphics[width=8.5cm]{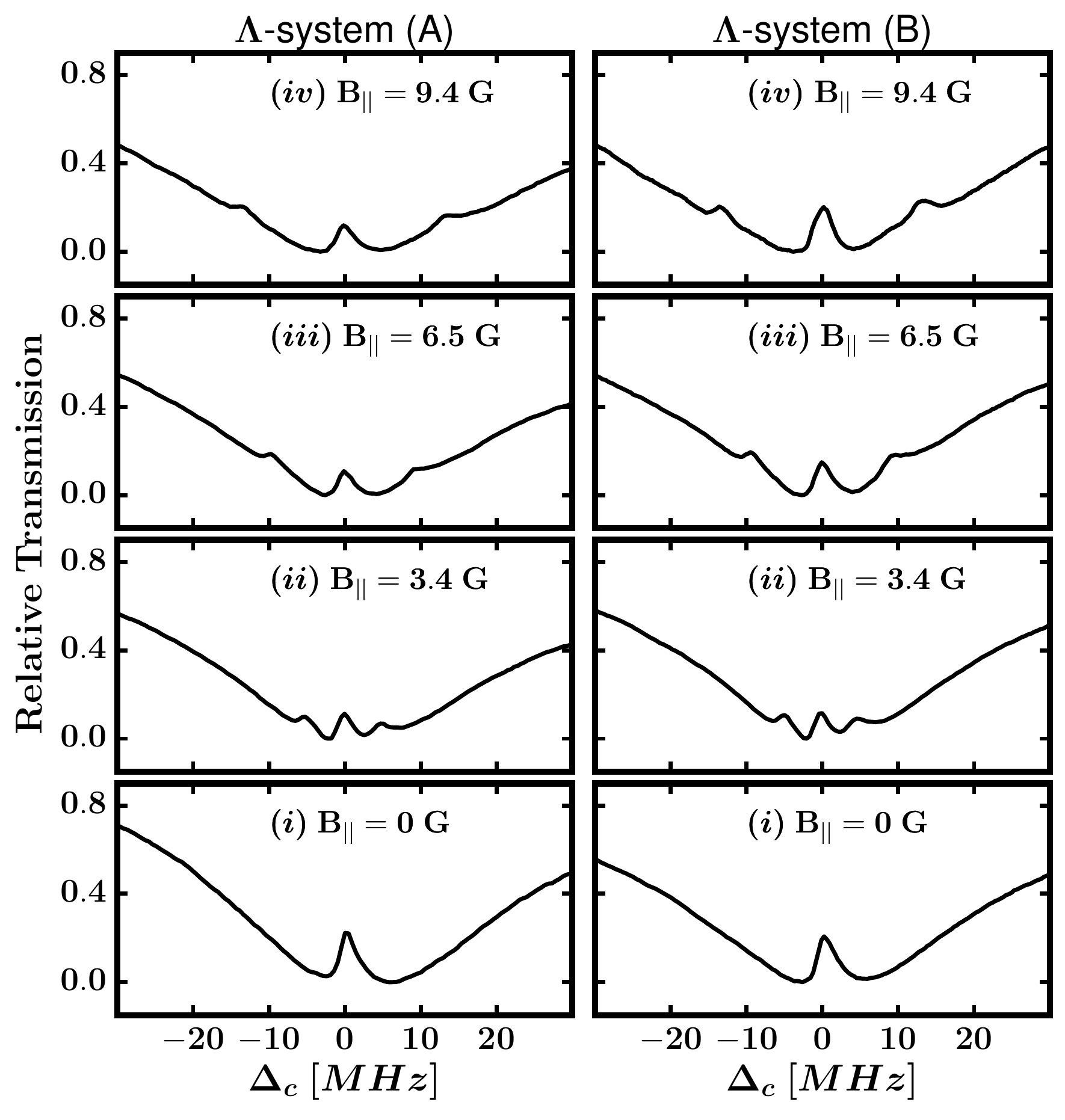}
\caption{ The relative transmission as a function of coupling beam detuning at different strength of the magnetic field ($B_{||}$) for both the $\Lambda$-systems. Here probe power is 0.08 mW and coupling power is 4 mW}.
\label{spectraB}
\end{figure}

Thereafter, experiments were performed by applying a longitudinal magnetic field $B_{||}$ to the vapor cell. The EIT signals for both the $\Lambda$-systems were recorded for different strengths of the magnetic field ($B_{||}$), while keeping the probe and coupling power fixed ($P_{p}=0.08$ mW and $P_{c}=4$ mW). The results of these observations are shown in Figure \ref{spectraB}. It is evident from this figure that EIT signals for both the systems split into three peaks in presence of the magnetic field. However, the central peak (among the three split peaks) for the system (B) is of higher strength than that for the system (A). The origin of the splitting of the EIT peaks is due to the removal of the degeneracy of the Zeeman hyperfine states ($m_F$  and $m_{F'}$) in presence of the applied magnetic field. The energy separation between two adjacent Zeeman sublevels is given by,
\begin{equation}
\Delta\left(|g_F|,B_{||}\right)=\frac{|g_{F}|\mu_{B}B_{||}}{h},
\label{eq}
\end{equation}
where $g_{F}$ is the hyperfine Land\'e g-factor, $\mu_{B}$ is Bohr magneton and $h$ is Planck's constant.

Each of the linearly polarized probe and coupling beams can be decomposed into two opposite circularly polarized beams with  $\sigma^{+}$ and $\sigma^{-}$ polarizations with respect to the direction of the magnetic field. Along with the Zeeman sublevels, these circularly polarized beams give rise to multiple $\Lambda$-subsystems within each $\Lambda$-system as per the transition selection rules $\Delta m_{F}=\pm 1$. The two-photon resonance condition in these $\Lambda$-subsystems should result in emergence of multiple EIT windows at different values of coupling beam detuning. It can be shown that there are only three possible values of coupling beam detuning for which the required two-photon resonance condition for EIT can be achieved. This can be explained as below.

Due to the different Lande g-factors associated with the ground and excited states (\textit{i.e.} $g_F$ values), the separation between their magnetic sublevels are different and are denoted as $\Delta_g$ (\textit{i.e.} $\Delta(|g_F|=1/2)$) and $\Delta_e$ (\textit{i.e.} $\Delta(|g_F|=2/3)$) respectively as shown in Figure \ref{lvl} (b). Now, a particular $\Lambda$-subsystem of the system (B) formed with the transitions $|F=1,m_{F}=1 \rangle \rightarrow |F'=2,m_{F'}=2 \rangle \leftarrow |F=2,m_{F}=1 \rangle$ is considered where the probe and coupling beam detunings can be determined in terms of $\Delta_g$ and $\Delta_e$ as $\Delta_{p} = \Delta_{g} + 2\Delta_{e}$ and $\Delta_{c}=-\Delta_{g}+2\Delta_{e}$ respectively. The two photon resonance condition becomes $\Delta_{p}-\Delta_{c}= 2\Delta_{g}$. Since the probe is kept resonant (\textit{i.e.} $\Delta_{p}=0$), the two photon resonance condition required for EIT can be achieved at $\Delta_{c}= -2\Delta_{g}$. A similar calculation for all the other $\Lambda$-subsystems results in three different values of coupling beam detuning $\Delta_{c}= 0,\pm 2\Delta_{g}$ for which the EIT resonance condition is satisfied.

\begin{figure}[t]
\centering
\includegraphics[width=8.5cm]{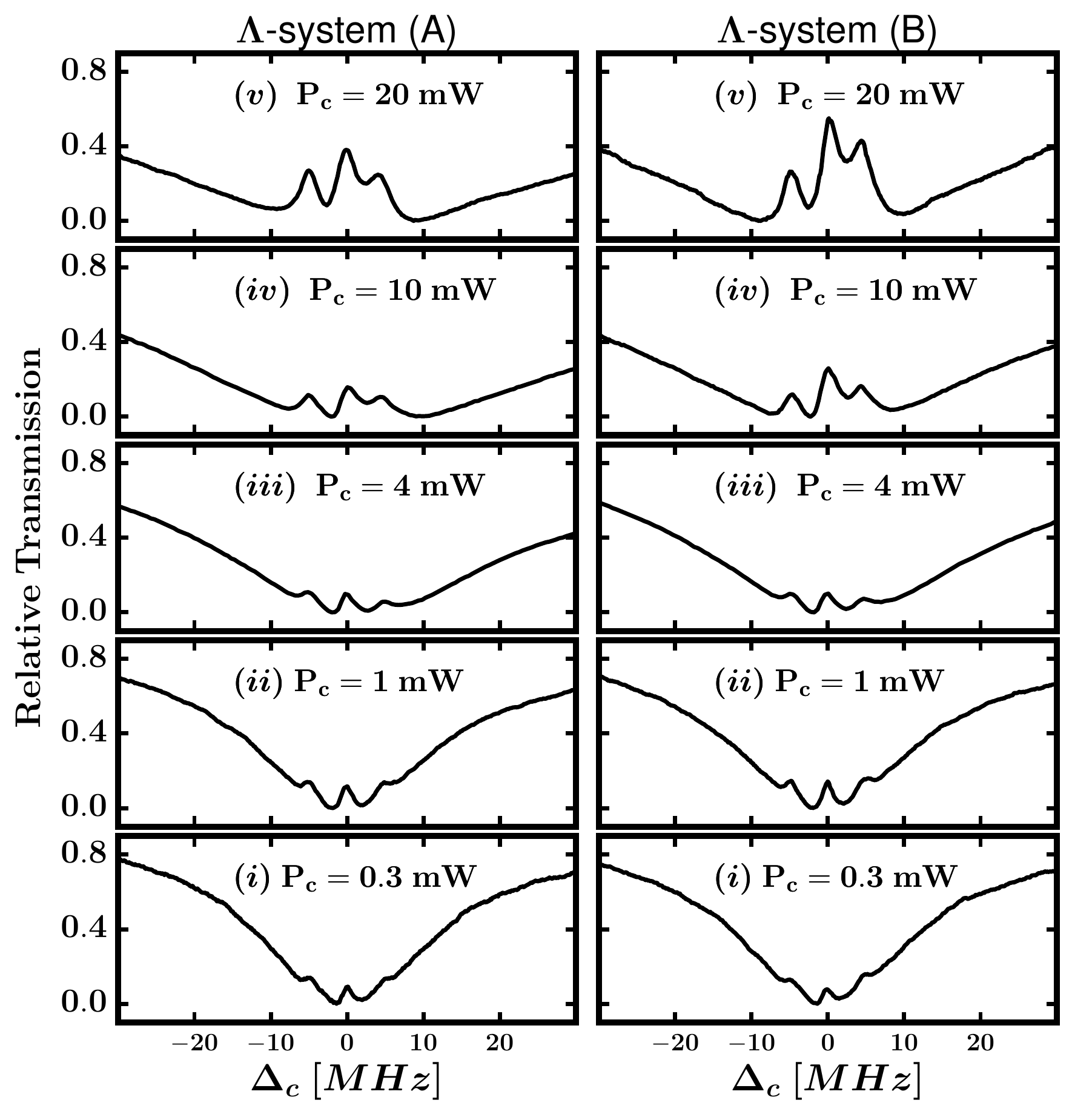}
\caption{ The relative transmission as a function of coupling beam detuning for both the $\Lambda$-systems in presence of longitudinal magnetic field $B_{||}=3.4\ G$. Plots \textit{(i)}-\textit{(v)} are for different coupling beam power at a fixed probe beam power of 0.08 mW}.
\label{spectraPcB}
\end{figure}

\begin{figure}[t]
\centering
\includegraphics[width=8.5cm]{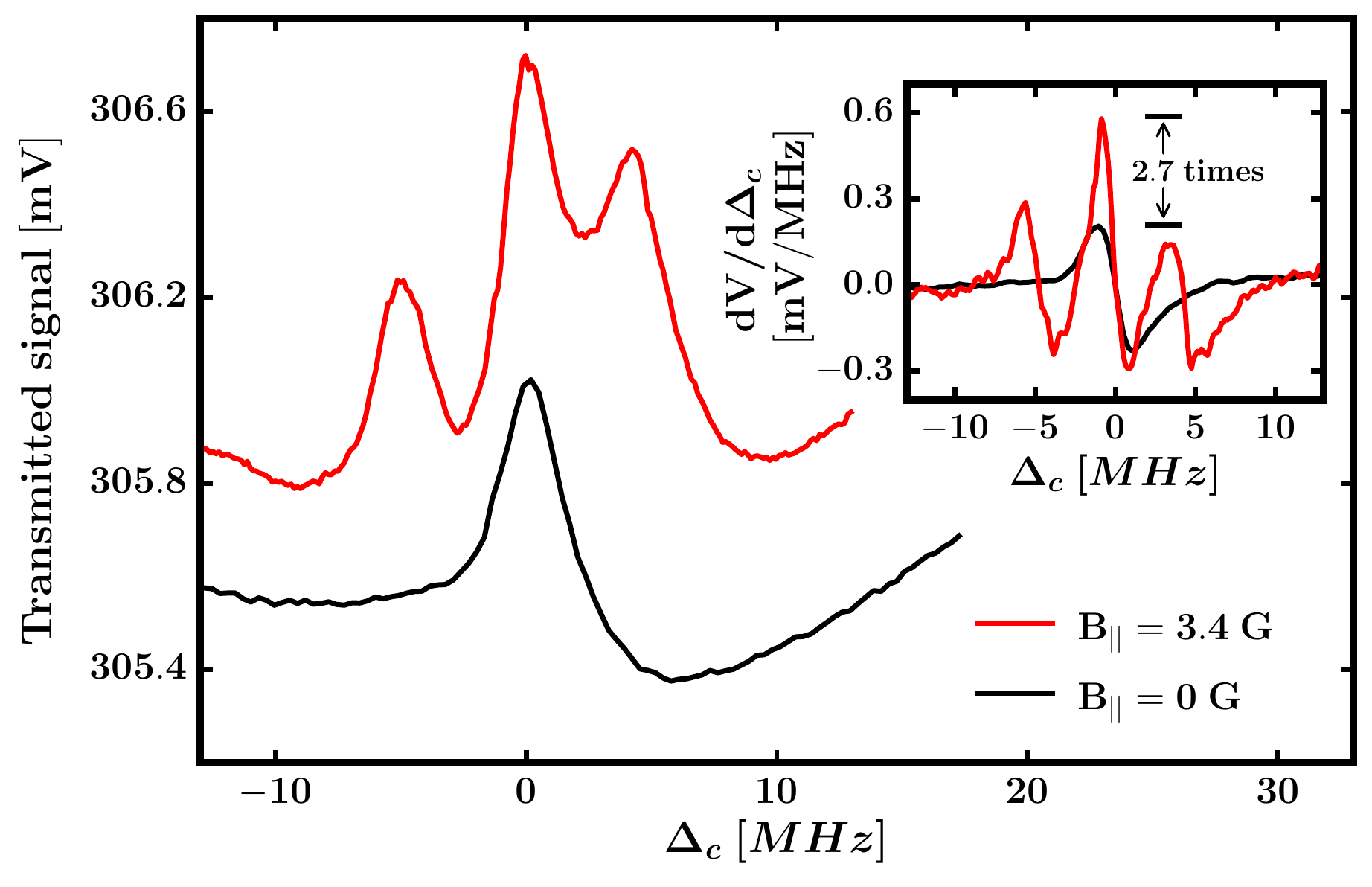}
\caption{(color online) The transmitted probe signal as a function of coupling beam detuning for $\Lambda$-system (B) in absence and in presence of the longitudinal magnetic field (\textit{i.e.} $B_{||}=0\ G$ (black curve) and $B_{||}=3.4\ G$ (red curve) respectively). The signals are for coupling beam power of 20 mW and probe beam power of 0.08 mW. The signal shown in black (lower curve) is shifted down by 10 mV for better visibility. The inset depicts the derivative of the transmitted signal with respect to the coupling beam detuning (\textit{i.e.} piecewise slope of the signal) as a function of coupling beam detuning. The slope in the presence of magnetic field is 2.7 times higher than the slope in absence of magnetic field}.
\label{slope}
\end{figure}
The behavior of these split EIT signals has also been studied by varying the coupling beam power, while keeping the magnetic field strength fixed at $B_{||}=$3.4 G and the probe beam power at $P_{p}=0.08$ mW. The results of this study are shown in Figure \ref{spectraPcB} for both the systems (A) and (B). It is evident from Figure \ref{spectraPcB} that the separation between the split-EIT peaks is independent of the coupling beam power whereas the strength of the EIT peaks increases by an increase in the coupling beam power. It can be noted that for coupling power greater than 4 mW, the central EIT peaks show higher amplitude with steeper slopes as compared to the EIT peaks observed in absence of the magnetic field. As an example, for a coupling power 20 mW in the system (B), both the signals are shown in Figure \ref{slope}. The slope of the central EIT peak in presence of magnetic field is $\sim$ 2.7 times higher than that of the EIT peak in absence of magnetic field which is clearly visible from the inset of Figure \ref{slope}. Because of the higher slope, the magnetic field induced split EIT signals may be a better choice for accurate frequency locking of the lasers. Along with this, side split peaks can also be utilized to lock a laser at frequencies controlled by the magnetic field strength.

\subsection{Numerical Simulations}
\label{sec:simulation}
To model the experimental results, all the hyperfine states of the $D_{2}$ line transition of $^{87}Rb$ atom were considered and the resulting six-level system is shown in Figure \ref{lvl}(a). Here the levels $|0 \rangle$ and $|1 \rangle$ represent two ground state hyperfine levels $F=1$ and $F=2$ and the levels $|2 \rangle$, $|3 \rangle$, $|4 \rangle$ and $|5 \rangle$ represent the excited state hyperfine levels $F'=0$, $F'=1$, $F'=2$ and $F'=3$ respectively. The $\Lambda$-systems (A) and (B) involve transitions $|0 \rangle \rightarrow |3 \rangle \leftarrow |1 \rangle$ and $|0 \rangle \rightarrow |4 \rangle \leftarrow |1 \rangle$ respectively. Therefore, as an example, the Hamiltonian for the $\Lambda$-system (B) with six-levels ($|0 \rangle$ to $|5 \rangle$) interacting with the probe and coupling fields after applying the rotating-wave approximation can be constructed as, 

\footnotesize 
\begin{equation}\label{eq:6Lhamiltonian}
H=\begin{bmatrix} 
\Delta_{p}-kv &0& &\Omega_{p}^{02}& &\Omega_{p}^{03}&  &\Omega_{p}^{04}& 0\\
0 &\Delta_{c}-kv& &0& &\Omega_{c}^{13}& &\Omega_{c}^{14}& \Omega_{c}^{15}\\
\Omega_{p}^{20} &0& &-2\delta_{42}& &0& &0& 0 \\
\Omega_{p}^{30} &\Omega_{c}^{31}& &0& &-2\delta_{43}& &0& 0 \\
\Omega_{p}^{40} &\Omega_{c}^{41}& &0& &0& &0& 0\\
0 &\Omega_{c}^{51}& &0& &0& &0& 2\delta_{45}   
\end{bmatrix},
\end{equation} 
\normalsize

where $\Delta_{p}$ and $\Delta_{c}$ are the frequency detunings of the probe and coupling beams respectively from their resonances. The Rabi frequency for the transitions $|i \rangle \rightarrow |j \rangle$ is denoted by $\Omega_{p,c}^{ij}$ where $p$ and $c$ denotes the probe and coupling transitions and the separation between the energy levels $|i \rangle$ and $|j \rangle$ is denoted as $\delta_{ij}$. The diagonal elements of $H$ arise due to the contribution from the atomic Hamiltonian and the off-diagonal terms manifest the interaction of an atom with the probe and coupling fields. The term $kv$ accounts for the thermal motion of the atom, where $v$ is the velocity of an atom and \textit{k} is the wave vector of the electromagnetic fields. The evolution of this atomic system, in terms of the density matrix ($\rho$), can be described by the Lindblad master equation

\begin{equation} \label{eq:rho}
\dot\rho=-\frac{i}{\hbar}[H,\rho]+L(\rho).
\end{equation}

\begin{figure}[t]
\centering
\includegraphics[width=8.5cm]{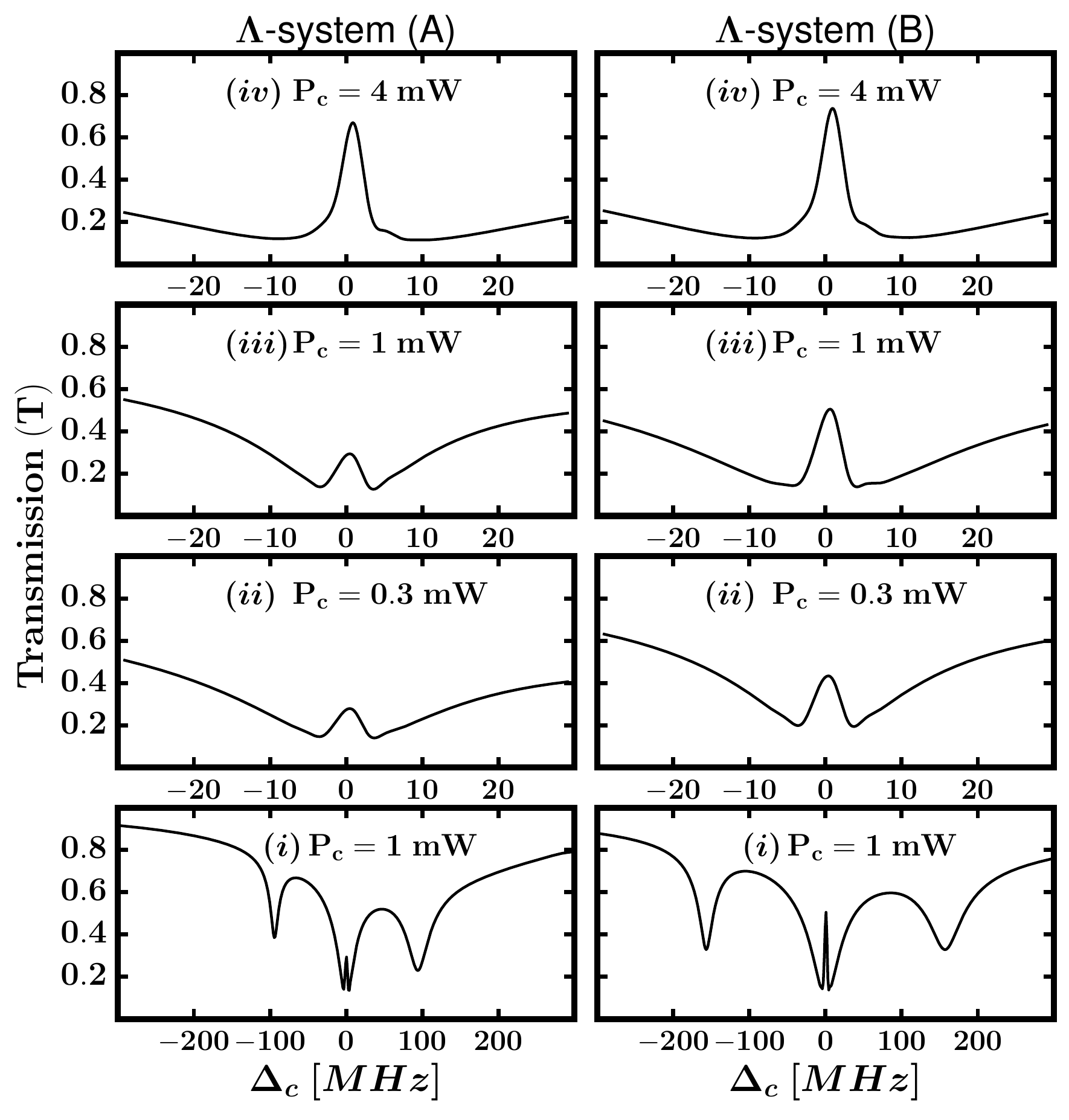}
\caption{The simulated transmission of probe beam as a function of coupling beam detuning for both the $\Lambda$-systems at different coupling beam power, in absence of magnetic field. Plot (\textit{i}) shows the probe transmission for a large range of coupling beam detuning, depicting the VSOP and EIT signals together. Plots (\textit{ii}) - (\textit{iv}) show magnified view of the EIT signal for different coupling power}.
\label{theory}
\end{figure}

Here L($\rho$) is the Lindblad super-operator which incorporates the effect of the spontaneous emissions in the system. This Lindblad master equation yields a set of thirty-six time dependent equations, and the steady state solution of these equations are obtained numerically. The imaginary parts of different coherences (\textit{i.e.} $Im(\rho_{02})$, $Im(\rho_{03})$ and $Im(\rho_{04})$) have been used to calculate the imaginary part of the linear susceptibility $Im(\chi^{(1)})$, which in turn provides the probe transmission. Due to the presence of a large number of atoms with a Maxwell Boltzmann velocity distribution ($W(v)$) in the room temperature Rb vapor cell, the imaginary part of the susceptibility was averaged over all the velocities \cite{Peters:2012} as, 

\begin{equation} \label{eq:chi}
Im(\chi^{(1)})=\int dv W(v) A \sum\limits_{j=2}^{4} \left(\frac{\mu_{0j}^{2}\times Im(\rho_{0j})}{\Omega_{p}^{0j}}\right),\\
\end{equation} 

where A=$\frac{2n}{\epsilon_{0} \hbar}$, $\mu_{0j}$ is the dipole moment between states $|0\rangle$ and $|j\rangle$, $\epsilon_{0}$ is vacuum permittivity, $\hbar$ is reduced Planck's constant and $n$ is number density of $Rb$ atoms inside the $Rb$ vapor cell. The transmission of the medium is given as, 

\begin{equation} \label{eq:T}
T=\exp[-\alpha l],
\end{equation} 

where $\alpha=k Im(\chi^{(1)})$ is absorption coefficient, \textit{k} is the wave vector of the electromagnetic field and \textit{l} is the length of the vapor cell. The values of  parameters n, l, $\Omega_p$, $\Omega_c$, $\delta_{ij}$ used in the simulations were kept same as experimental values as given in sections \ref{sec:expt} and \ref{sec:expt_result}. The results of the simulations are shown in Figure \ref{theory} for both the $\Lambda$-systems.

As evident from Figures \ref{theory} and \ref{vsop}, the results obtained from the numerical simulations are able to reproduce the experimentally observed EIT signal along with the velocity selective optical pumping absorption dips for both the $\Lambda$-systems. The simulation results also show the increase in the EIT signal strength as well as the asymmetry in line shape with the increase in coupling beam power. However, the simulated EIT signal is less asymmetric as compared to the experimentally observed signal. Incorporation of various factors like ground state relaxation rate, laser linewidth, effect of collisions, non-radiative decay of atoms within the hyperfine levels of the excited states, which are ignored in the simulations, may result in better agreement between the experimental and simulation results \cite{Figueroa:2006,Lu:1997,Ghosh:2009}. The numerical results in Figure \ref{theory} also proves that the system (B) exhibit EIT of better strength than that of the system (A) for a given coupling power.

\begin{figure}[t]
\centering
\includegraphics[width=8.5cm]{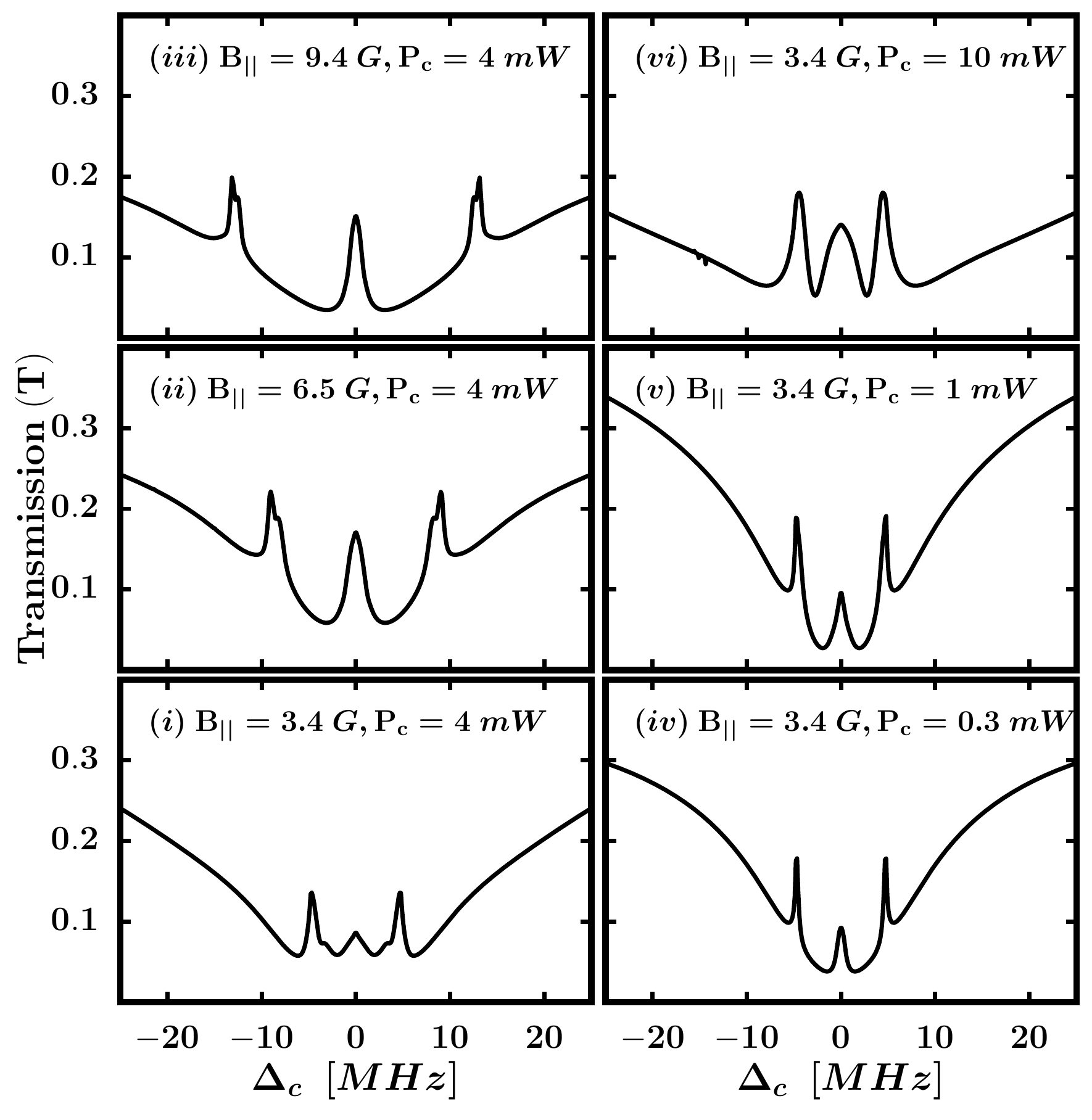}
\caption{The simulated transmission of probe beam as a function of coupling beam detuning for $\Lambda$-system (B) in presence of a magnetic field. Curves \textit{(i),(ii) and (iii)} show the variation in separation of split EIT peaks with the magnetic field strength for a fixed coupling beam power of 4 mW. Curves \textit{(iv),(v) and (vi)} show the variation in amplitude and line shape of split EIT peaks with the coupling beam power at a fixed magnetic field strength 3.4 G}.
\label{sim_B}
\end{figure}

In order to model the observed EIT signals in presence of a magnetic field for both the systems (A) and (B), the splitting of the hyperfine levels (\textit{i.e.} 11 sublevels for system (A) and 13 sublevels for system (B)) should also be considered. The Hamiltonian for $\Lambda$-system (B) with all the 13 Zeeman sublevels interacting with the applied probe and coupling beams can be written as,
\begin{multline} \label{eq:H:mf}
H=\hbar\sum\limits_{i=0}^{12}\omega_{ii} |i \rangle \langle i|+\frac{\hbar}{2}\sum\limits_{i=0 }^{i=2}\sum\limits_{\substack{j=8} }^{j=12} \Omega_p^{i,j} |i \rangle \langle j|+\\ \frac{\hbar}{2}\sum\limits_{i=3}^{i=7}\sum\limits_{\substack{j=8} }^{j=12}\Omega_c^{i,j} |i \rangle \langle j|+H.c.,
\end{multline}
where $i$ and $j$ correspond to different Zeeman hyperfine ground states ($m_{F}$) and excited states ($m_{F'}$) respectively and the corresponding transitions between $i^{th}$ and $j^{th}$ states follow the dipole selection rule $m_{F'}-m_{F}=\pm 1$ (Figure \ref{lvl}(b)). Here $\hbar \omega_{ii}$ represents the energy of $i^{th}$ state.

The Lindblad master equation (\ref{eq:rho}) was solved using the above Hamiltonian of equation (\ref{eq:H:mf}) to calculate the probe transmission in presence of the magnetic field. Numerically obtained values of probe transmission T as a function of coupling beam detuning for different values of magnetic field strength and coupling beam power are shown in Figure \ref{sim_B}. The variation in the peak separation with the applied magnetic field as obtained numerically (Figure \ref{sim_B} (i)-(iii)) are in good agreement with the experimental observations (Figure \ref{spectraB} (ii)-(iv) for $\Lambda$-system (B)). The occurrence of three split EIT peaks in the simulation results also agree with the analytical explanation provided in the earlier section.

In numerical simulations, the variation in the line shape and amplitude of split EIT peaks with the coupling beam power is shown in Figure \ref{sim_B} (curves (iv)-(vi)), for a fixed magnetic field strength 3.4 G. As the coupling power is increased, the simulated spectra show broadening in all the three EIT peaks. These results are also in qualitative agreement with the experimental observations. However, a better matching between the theoretical and experimental results may be obtained by incorporating various effects such as collisions \cite{Ghosh:2009}, spin exchange relaxations \cite{Shuker:2008}, etc in the theoretical model. Nonetheless, the presented numerical results provide an insight into the EIT phenomenon in these multi-level systems and can be employed to investigate similar other systems.

\section{CONCLUSION}
\label{sec:concl}

The two $\Lambda$-systems in $^{87}Rb$ atom, $|5^{2}S_{1/2} F=1\rangle \rightarrow |5^{2}P_{3/2} F'=1\rangle \leftarrow |5^{2}S_{1/2} F=2\rangle$  and $|5^{2}S_{1/2} F=1\rangle \rightarrow |5^{2}P_{3/2} F'=2\rangle \leftarrow |5^{2}S_{1/2} F=2\rangle$, denoted as system (A) and (B) respectively, have been investigated for EIT characteristics by varying the coupling beam and magnetic field strength. The experimentally observed results have been explained with our multi-level models for EIT phenomenon. Though, the observed linewidth of EIT signals is found to remain sub-natural even at higher coupling power for both the systems, the EIT signals for the system (B) have higher strength and more symmetric line shape than corresponding signals for system (A). At a higher coupling beam power and in presence of the magnetic field, the central EIT peak possess steeper slope than the peak observed in absence of the magnetic field. These signals with steeper slope can be extremely useful for tight laser frequency locking. Locking of a laser frequency at a frequency detuning controlled by the magnetic field can be achieved using the dependence of the separation between the split peaks on the magnetic field strength. The observed results indicate that the system (B) is more promising from  application point of view.

\section{ACKNOWLEDGMENTS}

We are thankful to S. Singh and V. Singh for their help during experiment. Charu Mishra is grateful for financial support from RRCAT, Indore under HBNI, Mumbai program.


\begin{thebibliography}{43}
\expandafter\ifx\csname natexlab\endcsname\relax\def\natexlab#1{#1}\fi
\expandafter\ifx\csname bibnamefont\endcsname\relax
  \def\bibnamefont#1{#1}\fi
\expandafter\ifx\csname bibfnamefont\endcsname\relax
  \def\bibfnamefont#1{#1}\fi
\expandafter\ifx\csname citenamefont\endcsname\relax
  \def\citenamefont#1{#1}\fi
\expandafter\ifx\csname url\endcsname\relax
  \def\url#1{\texttt{#1}}\fi
\expandafter\ifx\csname urlprefix\endcsname\relax\def\urlprefix{URL }\fi
\providecommand{\bibinfo}[2]{#2}
\providecommand{\eprint}[2][]{\url{#2}}

\bibitem[{\citenamefont{Harris et~al.}(1990)\citenamefont{Harris, Field, and
  Imamo\ifmmode~\breve{g}\else \u{g}\fi{}lu}}]{Harris:1990}
\bibinfo{author}{\bibfnamefont{S.~E.} \bibnamefont{Harris}},
  \bibinfo{author}{\bibfnamefont{J.~E.} \bibnamefont{Field}}, \bibnamefont{and}
  \bibinfo{author}{\bibfnamefont{A.}~\bibnamefont{Imamo\ifmmode~\breve{g}\else
  \u{g}\fi{}lu}}, \bibinfo{journal}{Phys. Rev. Lett.}
  \textbf{\bibinfo{volume}{64}}, \bibinfo{pages}{1107} (\bibinfo{year}{1990}),
  \urlprefix\url{http://link.aps.org/doi/10.1103/PhysRevLett.64.1107}.

\bibitem[{\citenamefont{Boller et~al.}(1991)\citenamefont{Boller,
  Imamo\ifmmode~\breve{g}\else \u{g}\fi{}lu, and Harris}}]{Harris:1st}
\bibinfo{author}{\bibfnamefont{K.-J.} \bibnamefont{Boller}},
  \bibinfo{author}{\bibfnamefont{A.}~\bibnamefont{Imamo\ifmmode~\breve{g}\else
  \u{g}\fi{}lu}}, \bibnamefont{and} \bibinfo{author}{\bibfnamefont{S.~E.}
  \bibnamefont{Harris}}, \bibinfo{journal}{Phys. Rev. Lett.}
  \textbf{\bibinfo{volume}{66}}, \bibinfo{pages}{2593} (\bibinfo{year}{1991}),
  \urlprefix\url{http://link.aps.org/doi/10.1103/PhysRevLett.66.2593}.

\bibitem[{\citenamefont{Ma et~al.}(2017)\citenamefont{Ma, Slattery, and
  Tang}}]{Ma:2017}
\bibinfo{author}{\bibfnamefont{L.}~\bibnamefont{Ma}},
  \bibinfo{author}{\bibfnamefont{O.}~\bibnamefont{Slattery}}, \bibnamefont{and}
  \bibinfo{author}{\bibfnamefont{X.}~\bibnamefont{Tang}},
  \bibinfo{journal}{Journal of Optics} \textbf{\bibinfo{volume}{19}},
  \bibinfo{pages}{043001} (\bibinfo{year}{2017}),
  \urlprefix\url{http://stacks.iop.org/2040-8986/19/i=4/a=043001}.

\bibitem[{\citenamefont{Clarke et~al.}(2001{\natexlab{a}})\citenamefont{Clarke,
  Chen, and van Wijngaarden}}]{Clarke:12:2001}
\bibinfo{author}{\bibfnamefont{J.}~\bibnamefont{Clarke}},
  \bibinfo{author}{\bibfnamefont{H.}~\bibnamefont{Chen}}, \bibnamefont{and}
  \bibinfo{author}{\bibfnamefont{W.~A.} \bibnamefont{van Wijngaarden}},
  \bibinfo{journal}{Appl. Opt.} \textbf{\bibinfo{volume}{40}},
  \bibinfo{pages}{2047} (\bibinfo{year}{2001}{\natexlab{a}}),
  \urlprefix\url{http://ao.osa.org/abstract.cfm?URI=ao-40-12-2047}.

\bibitem[{\citenamefont{Ray et~al.}(2013)\citenamefont{Ray, Sabir~Ali, and
  Chakrabarti}}]{Ray:2013}
\bibinfo{author}{\bibfnamefont{A.}~\bibnamefont{Ray}},
  \bibinfo{author}{\bibfnamefont{M.}~\bibnamefont{Sabir~Ali}},
  \bibnamefont{and}
  \bibinfo{author}{\bibfnamefont{A.}~\bibnamefont{Chakrabarti}},
  \bibinfo{journal}{The European Physical Journal D}
  \textbf{\bibinfo{volume}{67}}, \bibinfo{pages}{78} (\bibinfo{year}{2013}),
  ISSN \bibinfo{issn}{1434-6079},
  \urlprefix\url{"https://doi.org/10.1140/epjd/e2013-30653-1"}.

\bibitem[{\citenamefont{Kash et~al.}(1999)\citenamefont{Kash, Sautenkov,
  Zibrov, Hollberg, Welch, Lukin, Rostovtsev, Fry, and Scully}}]{Kash:1999}
\bibinfo{author}{\bibfnamefont{M.~M.} \bibnamefont{Kash}},
  \bibinfo{author}{\bibfnamefont{V.~A.} \bibnamefont{Sautenkov}},
  \bibinfo{author}{\bibfnamefont{A.~S.} \bibnamefont{Zibrov}},
  \bibinfo{author}{\bibfnamefont{L.}~\bibnamefont{Hollberg}},
  \bibinfo{author}{\bibfnamefont{G.~R.} \bibnamefont{Welch}},
  \bibinfo{author}{\bibfnamefont{M.~D.} \bibnamefont{Lukin}},
  \bibinfo{author}{\bibfnamefont{Y.}~\bibnamefont{Rostovtsev}},
  \bibinfo{author}{\bibfnamefont{E.~S.} \bibnamefont{Fry}}, \bibnamefont{and}
  \bibinfo{author}{\bibfnamefont{M.~O.} \bibnamefont{Scully}},
  \bibinfo{journal}{Phys. Rev. Lett.} \textbf{\bibinfo{volume}{82}},
  \bibinfo{pages}{5229} (\bibinfo{year}{1999}),
  \urlprefix\url{http://link.aps.org/doi/10.1103/PhysRevLett.82.5229}.

\bibitem[{\citenamefont{Kocharovskaya et~al.}(2001)\citenamefont{Kocharovskaya,
  Rostovtsev, and Scully}}]{Kocharovskaya:2001}
\bibinfo{author}{\bibfnamefont{O.}~\bibnamefont{Kocharovskaya}},
  \bibinfo{author}{\bibfnamefont{Y.}~\bibnamefont{Rostovtsev}},
  \bibnamefont{and} \bibinfo{author}{\bibfnamefont{M.~O.}
  \bibnamefont{Scully}}, \bibinfo{journal}{Phys. Rev. Lett.}
  \textbf{\bibinfo{volume}{86}}, \bibinfo{pages}{628} (\bibinfo{year}{2001}),
  \urlprefix\url{http://link.aps.org/doi/10.1103/PhysRevLett.86.628}.

\bibitem[{\citenamefont{Karpa and Weitz}(2006)}]{Karpa:2006}
\bibinfo{author}{\bibfnamefont{L.}~\bibnamefont{Karpa}} \bibnamefont{and}
  \bibinfo{author}{\bibfnamefont{M.}~\bibnamefont{Weitz}},
  \bibinfo{journal}{nature physics} \textbf{\bibinfo{volume}{2}},
  \bibinfo{pages}{332} (\bibinfo{year}{2006}).

\bibitem[{\citenamefont{Goldfarb et~al.}(2008)\citenamefont{Goldfarb, Ghosh,
  David, Ruggiero, Chanelière, Gouët, Gilles, Ghosh, and
  Bretenaker}}]{Goldfarb:2008}
\bibinfo{author}{\bibfnamefont{F.}~\bibnamefont{Goldfarb}},
  \bibinfo{author}{\bibfnamefont{J.}~\bibnamefont{Ghosh}},
  \bibinfo{author}{\bibfnamefont{M.}~\bibnamefont{David}},
  \bibinfo{author}{\bibfnamefont{J.}~\bibnamefont{Ruggiero}},
  \bibinfo{author}{\bibfnamefont{T.}~\bibnamefont{Chanelière}},
  \bibinfo{author}{\bibfnamefont{J.-L.~L.} \bibnamefont{Gouët}},
  \bibinfo{author}{\bibfnamefont{H.}~\bibnamefont{Gilles}},
  \bibinfo{author}{\bibfnamefont{R.}~\bibnamefont{Ghosh}}, \bibnamefont{and}
  \bibinfo{author}{\bibfnamefont{F.}~\bibnamefont{Bretenaker}},
  \bibinfo{journal}{EPL (Europhysics Letters)} \textbf{\bibinfo{volume}{82}},
  \bibinfo{pages}{54002} (\bibinfo{year}{2008}),
  \urlprefix\url{http://stacks.iop.org/0295-5075/82/i=5/a=54002}.

\bibitem[{\citenamefont{Alotaibi and Sanders}(2015)}]{Alotaibi:2015}
\bibinfo{author}{\bibfnamefont{H.~M.~M.} \bibnamefont{Alotaibi}}
  \bibnamefont{and} \bibinfo{author}{\bibfnamefont{B.~C.}
  \bibnamefont{Sanders}}, \bibinfo{journal}{Phys. Rev. A}
  \textbf{\bibinfo{volume}{91}}, \bibinfo{pages}{043817}
  (\bibinfo{year}{2015}),
  \urlprefix\url{http://link.aps.org/doi/10.1103/PhysRevA.91.043817}.

\bibitem[{\citenamefont{Bell et~al.}(2007)\citenamefont{Bell, Heywood, White,
  Close, and Scholten}}]{Bell:2007}
\bibinfo{author}{\bibfnamefont{S.~C.} \bibnamefont{Bell}},
  \bibinfo{author}{\bibfnamefont{D.~M.} \bibnamefont{Heywood}},
  \bibinfo{author}{\bibfnamefont{J.~D.} \bibnamefont{White}},
  \bibinfo{author}{\bibfnamefont{J.~D.} \bibnamefont{Close}}, \bibnamefont{and}
  \bibinfo{author}{\bibfnamefont{R.~E.} \bibnamefont{Scholten}},
  \bibinfo{journal}{Applied Physics Letters} \textbf{\bibinfo{volume}{90}},
  \bibinfo{pages}{171120} (\bibinfo{year}{2007}),
  \eprint{http://aip.scitation.org/doi/pdf/10.1063/1.2734471},
  \urlprefix\url{http://aip.scitation.org/doi/abs/10.1063/1.2734471}.

\bibitem[{\citenamefont{Krishna et~al.}(2005)\citenamefont{Krishna, Pandey,
  Wasan, and Natarajan}}]{Krishna:2005}
\bibinfo{author}{\bibfnamefont{A.}~\bibnamefont{Krishna}},
  \bibinfo{author}{\bibfnamefont{K.}~\bibnamefont{Pandey}},
  \bibinfo{author}{\bibfnamefont{A.}~\bibnamefont{Wasan}}, \bibnamefont{and}
  \bibinfo{author}{\bibfnamefont{V.}~\bibnamefont{Natarajan}},
  \bibinfo{journal}{EPL (Europhysics Letters)} \textbf{\bibinfo{volume}{72}},
  \bibinfo{pages}{221} (\bibinfo{year}{2005}),
  \urlprefix\url{http://stacks.iop.org/0295-5075/72/i=2/a=221}.

\bibitem[{\citenamefont{Kale et~al.}(2015)\citenamefont{Kale, Mishra, Tiwari,
  Singh, and Rawat}}]{Kale:2015}
\bibinfo{author}{\bibfnamefont{Y.~B.} \bibnamefont{Kale}},
  \bibinfo{author}{\bibfnamefont{S.~R.} \bibnamefont{Mishra}},
  \bibinfo{author}{\bibfnamefont{V.~B.} \bibnamefont{Tiwari}},
  \bibinfo{author}{\bibfnamefont{S.}~\bibnamefont{Singh}}, \bibnamefont{and}
  \bibinfo{author}{\bibfnamefont{H.~S.} \bibnamefont{Rawat}},
  \bibinfo{journal}{Phys. Rev. A} \textbf{\bibinfo{volume}{91}},
  \bibinfo{pages}{053852} (\bibinfo{year}{2015}),
  \urlprefix\url{http://link.aps.org/doi/10.1103/PhysRevA.91.053852}.

\bibitem[{\citenamefont{Guidry et~al.}(2017)\citenamefont{Guidry, Kuchina,
  Novikova, and Mikhailov}}]{Guidry:2017}
\bibinfo{author}{\bibfnamefont{M.~A.} \bibnamefont{Guidry}},
  \bibinfo{author}{\bibfnamefont{E.}~\bibnamefont{Kuchina}},
  \bibinfo{author}{\bibfnamefont{I.}~\bibnamefont{Novikova}}, \bibnamefont{and}
  \bibinfo{author}{\bibfnamefont{E.~E.} \bibnamefont{Mikhailov}},
  \bibinfo{journal}{J. Opt. Soc. Am. B} \textbf{\bibinfo{volume}{34}},
  \bibinfo{pages}{2244} (\bibinfo{year}{2017}),
  \urlprefix\url{http://josab.osa.org/abstract.cfm?URI=josab-34-10-2244}.

\bibitem[{\citenamefont{Beausoleil
  et~al.}(2004{\natexlab{a}})\citenamefont{Beausoleil, Munro, Rodrigues, and
  Spiller}}]{Beausoleil:eit}
\bibinfo{author}{\bibfnamefont{R.~G.} \bibnamefont{Beausoleil}},
  \bibinfo{author}{\bibfnamefont{W.~J.} \bibnamefont{Munro}},
  \bibinfo{author}{\bibfnamefont{D.~A.} \bibnamefont{Rodrigues}},
  \bibnamefont{and} \bibinfo{author}{\bibfnamefont{T.~P.}
  \bibnamefont{Spiller}}, \bibinfo{journal}{Journal of Modern Optics}
  \textbf{\bibinfo{volume}{51}}, \bibinfo{pages}{2441}
  (\bibinfo{year}{2004}{\natexlab{a}}),
  \eprint{http://dx.doi.org/10.1080/09500340408231802},
  \urlprefix\url{http://dx.doi.org/10.1080/09500340408231802}.

\bibitem[{\citenamefont{Beausoleil
  et~al.}(2004{\natexlab{b}})\citenamefont{Beausoleil, Munro, and
  Spiller}}]{Beausoleil:cpt}
\bibinfo{author}{\bibfnamefont{R.~G.} \bibnamefont{Beausoleil}},
  \bibinfo{author}{\bibfnamefont{W.~J.} \bibnamefont{Munro}}, \bibnamefont{and}
  \bibinfo{author}{\bibfnamefont{T.~P.} \bibnamefont{Spiller}},
  \bibinfo{journal}{Journal of Modern Optics} \textbf{\bibinfo{volume}{51}},
  \bibinfo{pages}{1559} (\bibinfo{year}{2004}{\natexlab{b}}),
  \eprint{http://dx.doi.org/10.1080/09500340408232474},
  \urlprefix\url{http://dx.doi.org/10.1080/09500340408232474}.

\bibitem[{\citenamefont{Lee et~al.}(1998)\citenamefont{Lee, Fleischhauer, and
  Scully}}]{Lee:1998}
\bibinfo{author}{\bibfnamefont{H.}~\bibnamefont{Lee}},
  \bibinfo{author}{\bibfnamefont{M.}~\bibnamefont{Fleischhauer}},
  \bibnamefont{and} \bibinfo{author}{\bibfnamefont{M.~O.}
  \bibnamefont{Scully}}, \bibinfo{journal}{Phys. Rev. A}
  \textbf{\bibinfo{volume}{58}}, \bibinfo{pages}{2587} (\bibinfo{year}{1998}),
  \urlprefix\url{http://link.aps.org/doi/10.1103/PhysRevA.58.2587}.

\bibitem[{\citenamefont{Fleischhauer and Scully}(1994)}]{Fleischhauer:1994}
\bibinfo{author}{\bibfnamefont{M.}~\bibnamefont{Fleischhauer}}
  \bibnamefont{and} \bibinfo{author}{\bibfnamefont{M.~O.}
  \bibnamefont{Scully}}, \bibinfo{journal}{Phys. Rev. A}
  \textbf{\bibinfo{volume}{49}}, \bibinfo{pages}{1973} (\bibinfo{year}{1994}),
  \urlprefix\url{http://link.aps.org/doi/10.1103/PhysRevA.49.1973}.

\bibitem[{\citenamefont{Yudin et~al.}(2010)\citenamefont{Yudin, Taichenachev,
  Dudin, Velichansky, Zibrov, and Zibrov}}]{Yudin:2010}
\bibinfo{author}{\bibfnamefont{V.~I.} \bibnamefont{Yudin}},
  \bibinfo{author}{\bibfnamefont{A.~V.} \bibnamefont{Taichenachev}},
  \bibinfo{author}{\bibfnamefont{Y.~O.} \bibnamefont{Dudin}},
  \bibinfo{author}{\bibfnamefont{V.~L.} \bibnamefont{Velichansky}},
  \bibinfo{author}{\bibfnamefont{A.~S.} \bibnamefont{Zibrov}},
  \bibnamefont{and} \bibinfo{author}{\bibfnamefont{S.~A.}
  \bibnamefont{Zibrov}}, \bibinfo{journal}{Phys. Rev. A}
  \textbf{\bibinfo{volume}{82}}, \bibinfo{pages}{033807}
  (\bibinfo{year}{2010}),
  \urlprefix\url{http://link.aps.org/doi/10.1103/PhysRevA.82.033807}.

\bibitem[{\citenamefont{Cox et~al.}(2011)\citenamefont{Cox, Yudin,
  Taichenachev, Novikova, and Mikhailov}}]{Cox:2011}
\bibinfo{author}{\bibfnamefont{K.}~\bibnamefont{Cox}},
  \bibinfo{author}{\bibfnamefont{V.~I.} \bibnamefont{Yudin}},
  \bibinfo{author}{\bibfnamefont{A.~V.} \bibnamefont{Taichenachev}},
  \bibinfo{author}{\bibfnamefont{I.}~\bibnamefont{Novikova}}, \bibnamefont{and}
  \bibinfo{author}{\bibfnamefont{E.~E.} \bibnamefont{Mikhailov}},
  \bibinfo{journal}{Phys. Rev. A} \textbf{\bibinfo{volume}{83}},
  \bibinfo{pages}{015801} (\bibinfo{year}{2011}),
  \urlprefix\url{https://link.aps.org/doi/10.1103/PhysRevA.83.015801}.

\bibitem[{\citenamefont{Margalit et~al.}(2013)\citenamefont{Margalit,
  Rosenbluh, and Wilson-Gordon}}]{Margalit:2013}
\bibinfo{author}{\bibfnamefont{L.}~\bibnamefont{Margalit}},
  \bibinfo{author}{\bibfnamefont{M.}~\bibnamefont{Rosenbluh}},
  \bibnamefont{and} \bibinfo{author}{\bibfnamefont{A.~D.}
  \bibnamefont{Wilson-Gordon}}, \bibinfo{journal}{Phys. Rev. A}
  \textbf{\bibinfo{volume}{87}}, \bibinfo{pages}{033808}
  (\bibinfo{year}{2013}),
  \urlprefix\url{http://link.aps.org/doi/10.1103/PhysRevA.87.033808}.

\bibitem[{\citenamefont{Cheng et~al.}(2017)\citenamefont{Cheng, Wang, Zhang,
  Xin, Luo, and Liu}}]{Cheng:2017}
\bibinfo{author}{\bibfnamefont{H.}~\bibnamefont{Cheng}},
  \bibinfo{author}{\bibfnamefont{H.-M.} \bibnamefont{Wang}},
  \bibinfo{author}{\bibfnamefont{S.-S.} \bibnamefont{Zhang}},
  \bibinfo{author}{\bibfnamefont{P.-P.} \bibnamefont{Xin}},
  \bibinfo{author}{\bibfnamefont{J.}~\bibnamefont{Luo}}, \bibnamefont{and}
  \bibinfo{author}{\bibfnamefont{H.-P.} \bibnamefont{Liu}},
  \bibinfo{journal}{Journal of Physics B: Atomic, Molecular and Optical
  Physics} \textbf{\bibinfo{volume}{50}}, \bibinfo{pages}{095401}
  (\bibinfo{year}{2017}),
  \urlprefix\url{http://stacks.iop.org/0953-4075/50/i=9/a=095401}.

\bibitem[{\citenamefont{dos Santos and Tabosa}(2006)}]{Santos:2006}
\bibinfo{author}{\bibfnamefont{F.~B.~M.} \bibnamefont{dos Santos}}
  \bibnamefont{and} \bibinfo{author}{\bibfnamefont{J.~W.~R.}
  \bibnamefont{Tabosa}}, \bibinfo{journal}{Phys. Rev. A}
  \textbf{\bibinfo{volume}{73}}, \bibinfo{pages}{023422}
  (\bibinfo{year}{2006}),
  \urlprefix\url{https://link.aps.org/doi/10.1103/PhysRevA.73.023422}.

\bibitem[{\citenamefont{Moon et~al.}(2005)\citenamefont{Moon, Lee, and
  Kim}}]{Moon:2005}
\bibinfo{author}{\bibfnamefont{H.~S.} \bibnamefont{Moon}},
  \bibinfo{author}{\bibfnamefont{L.}~\bibnamefont{Lee}}, \bibnamefont{and}
  \bibinfo{author}{\bibfnamefont{J.~B.} \bibnamefont{Kim}},
  \bibinfo{journal}{J. Opt. Soc. Am. B} \textbf{\bibinfo{volume}{22}},
  \bibinfo{pages}{2529} (\bibinfo{year}{2005}),
  \urlprefix\url{http://josab.osa.org/abstract.cfm?URI=josab-22-12-2529}.

\bibitem[{\citenamefont{Li and Xiao}(1995)}]{Li:1995}
\bibinfo{author}{\bibfnamefont{Y.-q.} \bibnamefont{Li}} \bibnamefont{and}
  \bibinfo{author}{\bibfnamefont{M.}~\bibnamefont{Xiao}},
  \bibinfo{journal}{Phys. Rev. A} \textbf{\bibinfo{volume}{51}},
  \bibinfo{pages}{R2703} (\bibinfo{year}{1995}),
  \urlprefix\url{http://link.aps.org/doi/10.1103/PhysRevA.51.R2703}.

\bibitem[{\citenamefont{Ying et~al.}(2014)\citenamefont{Ying, Niu, Chen, Cai,
  Qu, and Gong}}]{Kang:2014}
\bibinfo{author}{\bibfnamefont{K.}~\bibnamefont{Ying}},
  \bibinfo{author}{\bibfnamefont{Y.}~\bibnamefont{Niu}},
  \bibinfo{author}{\bibfnamefont{D.}~\bibnamefont{Chen}},
  \bibinfo{author}{\bibfnamefont{H.}~\bibnamefont{Cai}},
  \bibinfo{author}{\bibfnamefont{R.}~\bibnamefont{Qu}}, \bibnamefont{and}
  \bibinfo{author}{\bibfnamefont{S.}~\bibnamefont{Gong}},
  \bibinfo{journal}{Journal of Modern Optics} \textbf{\bibinfo{volume}{61}},
  \bibinfo{pages}{631} (\bibinfo{year}{2014}),
  \eprint{http://dx.doi.org/10.1080/09500340.2014.904019},
  \urlprefix\url{http://dx.doi.org/10.1080/09500340.2014.904019}.

\bibitem[{\citenamefont{{Fulton} et~al.}(1995)\citenamefont{{Fulton},
  {Moseley}, {Shepherd}, {Sinclair}, and {Dunn}}}]{Fulton:116:1995}
\bibinfo{author}{\bibfnamefont{D.~J.} \bibnamefont{{Fulton}}},
  \bibinfo{author}{\bibfnamefont{R.~R.} \bibnamefont{{Moseley}}},
  \bibinfo{author}{\bibfnamefont{S.}~\bibnamefont{{Shepherd}}},
  \bibinfo{author}{\bibfnamefont{B.~D.} \bibnamefont{{Sinclair}}},
  \bibnamefont{and} \bibinfo{author}{\bibfnamefont{M.~H.}
  \bibnamefont{{Dunn}}}, \bibinfo{journal}{Optics Communications}
  \textbf{\bibinfo{volume}{116}}, \bibinfo{pages}{231} (\bibinfo{year}{1995}).

\bibitem[{\citenamefont{Fulton et~al.}(1995)\citenamefont{Fulton, Shepherd,
  Moseley, Sinclair, and Dunn}}]{Fulton:52:1995}
\bibinfo{author}{\bibfnamefont{D.~J.} \bibnamefont{Fulton}},
  \bibinfo{author}{\bibfnamefont{S.}~\bibnamefont{Shepherd}},
  \bibinfo{author}{\bibfnamefont{R.~R.} \bibnamefont{Moseley}},
  \bibinfo{author}{\bibfnamefont{B.~D.} \bibnamefont{Sinclair}},
  \bibnamefont{and} \bibinfo{author}{\bibfnamefont{M.~H.} \bibnamefont{Dunn}},
  \bibinfo{journal}{Phys. Rev. A} \textbf{\bibinfo{volume}{52}},
  \bibinfo{pages}{2302} (\bibinfo{year}{1995}),
  \urlprefix\url{http://link.aps.org/doi/10.1103/PhysRevA.52.2302}.

\bibitem[{\citenamefont{Clarke et~al.}(2001{\natexlab{b}})\citenamefont{Clarke,
  van Wijngaarden, and Chen}}]{Clarke:64:2001}
\bibinfo{author}{\bibfnamefont{J.~J.} \bibnamefont{Clarke}},
  \bibinfo{author}{\bibfnamefont{W.~A.} \bibnamefont{van Wijngaarden}},
  \bibnamefont{and} \bibinfo{author}{\bibfnamefont{H.}~\bibnamefont{Chen}},
  \bibinfo{journal}{Phys. Rev. A} \textbf{\bibinfo{volume}{64}},
  \bibinfo{pages}{023818} (\bibinfo{year}{2001}{\natexlab{b}}),
  \urlprefix\url{http://link.aps.org/doi/10.1103/PhysRevA.64.023818}.

\bibitem[{\citenamefont{Yan et~al.}(2001)\citenamefont{Yan, Rickey, and
  Zhu}}]{Yan:2001}
\bibinfo{author}{\bibfnamefont{M.}~\bibnamefont{Yan}},
  \bibinfo{author}{\bibfnamefont{E.~G.} \bibnamefont{Rickey}},
  \bibnamefont{and} \bibinfo{author}{\bibfnamefont{Y.}~\bibnamefont{Zhu}},
  \bibinfo{journal}{J. Opt. Soc. Am. B} \textbf{\bibinfo{volume}{18}},
  \bibinfo{pages}{1057} (\bibinfo{year}{2001}),
  \urlprefix\url{http://josab.osa.org/abstract.cfm?URI=josab-18-8-1057}.

\bibitem[{\citenamefont{Tiwari et~al.}(2010)\citenamefont{Tiwari, Singh, Rawat,
  Singh, and Mehendale}}]{VBT:2010}
\bibinfo{author}{\bibfnamefont{V.~B.} \bibnamefont{Tiwari}},
  \bibinfo{author}{\bibfnamefont{S.}~\bibnamefont{Singh}},
  \bibinfo{author}{\bibfnamefont{H.~S.} \bibnamefont{Rawat}},
  \bibinfo{author}{\bibfnamefont{M.~P.} \bibnamefont{Singh}}, \bibnamefont{and}
  \bibinfo{author}{\bibfnamefont{S.~C.} \bibnamefont{Mehendale}},
  \bibinfo{journal}{Journal of Physics B: Atomic, Molecular and Optical
  Physics} \textbf{\bibinfo{volume}{43}}, \bibinfo{pages}{095503}
  (\bibinfo{year}{2010}),
  \urlprefix\url{http://stacks.iop.org/0953-4075/43/i=9/a=095503}.

\bibitem[{\citenamefont{Jiang et~al.}(2016)\citenamefont{Jiang, Zhang, and
  Wang}}]{Xiaojun:2016}
\bibinfo{author}{\bibfnamefont{X.}~\bibnamefont{Jiang}},
  \bibinfo{author}{\bibfnamefont{H.}~\bibnamefont{Zhang}}, \bibnamefont{and}
  \bibinfo{author}{\bibfnamefont{Y.}~\bibnamefont{Wang}},
  \bibinfo{journal}{Chinese Physics B} \textbf{\bibinfo{volume}{25}},
  \bibinfo{pages}{034204} (\bibinfo{year}{2016}),
  \urlprefix\url{http://stacks.iop.org/1674-1056/25/i=3/a=034204}.

\bibitem[{\citenamefont{Carvalho et~al.}(2004)\citenamefont{Carvalho,
  de~Araujo, and Tabosa}}]{carvalho}
\bibinfo{author}{\bibfnamefont{P.~R.~S.} \bibnamefont{Carvalho}},
  \bibinfo{author}{\bibfnamefont{L.~E.~E.} \bibnamefont{de~Araujo}},
  \bibnamefont{and} \bibinfo{author}{\bibfnamefont{J.~W.~R.}
  \bibnamefont{Tabosa}}, \bibinfo{journal}{Phys. Rev. A}
  \textbf{\bibinfo{volume}{70}}, \bibinfo{pages}{063818}
  (\bibinfo{year}{2004}),
  \urlprefix\url{http://link.aps.org/doi/10.1103/PhysRevA.70.063818}.

\bibitem[{\citenamefont{Iftiquar and Natarajan}(2009)}]{Iftiquar:2009}
\bibinfo{author}{\bibfnamefont{S.~M.} \bibnamefont{Iftiquar}} \bibnamefont{and}
  \bibinfo{author}{\bibfnamefont{V.}~\bibnamefont{Natarajan}},
  \bibinfo{journal}{Phys. Rev. A} \textbf{\bibinfo{volume}{79}},
  \bibinfo{pages}{013808} (\bibinfo{year}{2009}),
  \urlprefix\url{http://link.aps.org/doi/10.1103/PhysRevA.79.013808}.

\bibitem[{\citenamefont{Chakrabarti et~al.}(2005)\citenamefont{Chakrabarti,
  Pradhan, Ray, and Ghosh}}]{Chakrabarti:2005}
\bibinfo{author}{\bibfnamefont{S.}~\bibnamefont{Chakrabarti}},
  \bibinfo{author}{\bibfnamefont{A.}~\bibnamefont{Pradhan}},
  \bibinfo{author}{\bibfnamefont{B.}~\bibnamefont{Ray}}, \bibnamefont{and}
  \bibinfo{author}{\bibfnamefont{P.~N.} \bibnamefont{Ghosh}},
  \bibinfo{journal}{Journal of Physics B: Atomic, Molecular and Optical
  Physics} \textbf{\bibinfo{volume}{38}}, \bibinfo{pages}{4321}
  (\bibinfo{year}{2005}),
  \urlprefix\url{http://stacks.iop.org/0953-4075/38/i=23/a=013}.

\bibitem[{\citenamefont{Hossain et~al.}(2011)\citenamefont{Hossain, Mitra,
  Poddar, Chaudhuri, Ray, and Ghosh}}]{Hossain:2011}
\bibinfo{author}{\bibfnamefont{M.~M.} \bibnamefont{Hossain}},
  \bibinfo{author}{\bibfnamefont{S.}~\bibnamefont{Mitra}},
  \bibinfo{author}{\bibfnamefont{P.}~\bibnamefont{Poddar}},
  \bibinfo{author}{\bibfnamefont{C.}~\bibnamefont{Chaudhuri}},
  \bibinfo{author}{\bibfnamefont{B.}~\bibnamefont{Ray}}, \bibnamefont{and}
  \bibinfo{author}{\bibfnamefont{P.~N.} \bibnamefont{Ghosh}},
  \bibinfo{journal}{Journal of Physics B: Atomic, Molecular and Optical
  Physics} \textbf{\bibinfo{volume}{44}}, \bibinfo{pages}{115501}
  (\bibinfo{year}{2011}),
  \urlprefix\url{http://stacks.iop.org/0953-4075/44/i=11/a=115501}.

\bibitem[{\citenamefont{Bharti and Wasan}(2012)}]{Bharti:2012}
\bibinfo{author}{\bibfnamefont{V.}~\bibnamefont{Bharti}} \bibnamefont{and}
  \bibinfo{author}{\bibfnamefont{A.}~\bibnamefont{Wasan}},
  \bibinfo{journal}{Journal of Physics B: Atomic, Molecular and Optical
  Physics} \textbf{\bibinfo{volume}{45}}, \bibinfo{pages}{185501}
  (\bibinfo{year}{2012}),
  \urlprefix\url{http://stacks.iop.org/0953-4075/45/i=18/a=185501}.

\bibitem[{\citenamefont{{Chen, Zhuo Ren} and {Su, Xue Mei}}(2013)}]{Chen:2013}
\bibinfo{author}{\bibnamefont{{Chen, Zhuo Ren}}} \bibnamefont{and}
  \bibinfo{author}{\bibnamefont{{Su, Xue Mei}}}, \bibinfo{journal}{Eur. Phys.
  J. D} \textbf{\bibinfo{volume}{67}}, \bibinfo{pages}{138}
  (\bibinfo{year}{2013}),
  \urlprefix\url{https://doi.org/10.1140/epjd/e2013-40056-y}.

\bibitem[{\citenamefont{Peters et~al.}(2012)\citenamefont{Peters, Wittrock,
  Blatt, Halfmann, and Yatsenko}}]{Peters:2012}
\bibinfo{author}{\bibfnamefont{T.}~\bibnamefont{Peters}},
  \bibinfo{author}{\bibfnamefont{B.}~\bibnamefont{Wittrock}},
  \bibinfo{author}{\bibfnamefont{F.}~\bibnamefont{Blatt}},
  \bibinfo{author}{\bibfnamefont{T.}~\bibnamefont{Halfmann}}, \bibnamefont{and}
  \bibinfo{author}{\bibfnamefont{L.~P.} \bibnamefont{Yatsenko}},
  \bibinfo{journal}{Phys. Rev. A} \textbf{\bibinfo{volume}{85}},
  \bibinfo{pages}{063416} (\bibinfo{year}{2012}),
  \urlprefix\url{http://link.aps.org/doi/10.1103/PhysRevA.85.063416}.

\bibitem[{\citenamefont{Figueroa et~al.}(2006)\citenamefont{Figueroa, Vewinger,
  Appel, and Lvovsky}}]{Figueroa:2006}
\bibinfo{author}{\bibfnamefont{E.}~\bibnamefont{Figueroa}},
  \bibinfo{author}{\bibfnamefont{F.}~\bibnamefont{Vewinger}},
  \bibinfo{author}{\bibfnamefont{J.}~\bibnamefont{Appel}}, \bibnamefont{and}
  \bibinfo{author}{\bibfnamefont{A.~I.} \bibnamefont{Lvovsky}},
  \bibinfo{journal}{Opt. Lett.} \textbf{\bibinfo{volume}{31}},
  \bibinfo{pages}{2625} (\bibinfo{year}{2006}),
  \urlprefix\url{http://ol.osa.org/abstract.cfm?URI=ol-31-17-2625}.

\bibitem[{\citenamefont{L\"u et~al.}(1997)\citenamefont{L\"u, Burkett, and
  Xiao}}]{Lu:1997}
\bibinfo{author}{\bibfnamefont{B.}~\bibnamefont{L\"u}},
  \bibinfo{author}{\bibfnamefont{W.~H.} \bibnamefont{Burkett}},
  \bibnamefont{and} \bibinfo{author}{\bibfnamefont{M.}~\bibnamefont{Xiao}},
  \bibinfo{journal}{Phys. Rev. A} \textbf{\bibinfo{volume}{56}},
  \bibinfo{pages}{976} (\bibinfo{year}{1997}),
  \urlprefix\url{https://link.aps.org/doi/10.1103/PhysRevA.56.976}.

\bibitem[{\citenamefont{Ghosh et~al.}(2009)\citenamefont{Ghosh, Ghosh,
  Goldfarb, Le~Gou\"et, and Bretenaker}}]{Ghosh:2009}
\bibinfo{author}{\bibfnamefont{J.}~\bibnamefont{Ghosh}},
  \bibinfo{author}{\bibfnamefont{R.}~\bibnamefont{Ghosh}},
  \bibinfo{author}{\bibfnamefont{F.}~\bibnamefont{Goldfarb}},
  \bibinfo{author}{\bibfnamefont{J.-L.} \bibnamefont{Le~Gou\"et}},
  \bibnamefont{and}
  \bibinfo{author}{\bibfnamefont{F.}~\bibnamefont{Bretenaker}},
  \bibinfo{journal}{Phys. Rev. A} \textbf{\bibinfo{volume}{80}},
  \bibinfo{pages}{023817} (\bibinfo{year}{2009}),
  \urlprefix\url{https://link.aps.org/doi/10.1103/PhysRevA.80.023817}.

\bibitem[{\citenamefont{Shuker et~al.}(2008)\citenamefont{Shuker, Firstenberg,
  Sagi, Ben-kish, Davidson, and Ron}}]{Shuker:2008}
\bibinfo{author}{\bibfnamefont{M.}~\bibnamefont{Shuker}},
  \bibinfo{author}{\bibfnamefont{O.}~\bibnamefont{Firstenberg}},
  \bibinfo{author}{\bibfnamefont{Y.}~\bibnamefont{Sagi}},
  \bibinfo{author}{\bibfnamefont{A.}~\bibnamefont{Ben-kish}},
  \bibinfo{author}{\bibfnamefont{N.}~\bibnamefont{Davidson}}, \bibnamefont{and}
  \bibinfo{author}{\bibfnamefont{A.}~\bibnamefont{Ron}},
  \bibinfo{journal}{Phys. Rev. A} \textbf{\bibinfo{volume}{78}},
  \bibinfo{pages}{063818} (\bibinfo{year}{2008}),
  \urlprefix\url{https://link.aps.org/doi/10.1103/PhysRevA.78.063818}.

\end{thebibliography}

\end{document}